\documentclass[letterpaper, 10 pt, conference]{ieeeconf}  
                                                          \usepackage{mathrsfs}
\usepackage{amsmath}
\usepackage{amssymb}
\newtheorem{problem}{Problem}

\usepackage{mathrsfs}
\usepackage{amsmath}
\usepackage{enumerate}
\usepackage{marvosym}

\newcommand{\rank}{\text{rank}}

\newcommand{\tr}{\text{{Tr}}}

\newcommand{\diag}{\text{{diag}}}

\def\onesvec{{\bf 1}}
\def\diag{\text{{diag}}}

\IEEEoverridecommandlockouts                              
   \usepackage{graphicx}                                                       
\title{\LARGE \bf
Model Predictive Control of Collinear Coulomb Spacecraft Formations
}
\author{Adam M. Tahir 
\thanks{A. M. Tahir is with the William E. Boeing Department of Aeronautics and Astronautics, University of Washington, Seattle, WA, USA. Email: 
       {{\tt\small aerotahir@gmail.com}.}
}}

\begin{document}

\maketitle
\thispagestyle{empty}
\pagestyle{empty}

\begin{abstract}
A model predictive control scheme to stabilize desired configurations of collinear Coulomb spacecraft formations is derived in this paper. The nonlinearities of the dynamics with respect to the input make this problem difficult to solve, computationally. It is shown that the nonlinearities in the input lead to a finite horizon optimization problem which is a nonconvex quadratically-constrained quadratic program (QCQP). A convex relaxation of the nonconvex QCQP is therefore derived which can be solved quickly using a convex optimization solver. A simulation of a four spacecraft formation is provided which demonstrates why optimizing over a prediction horizon is a prudent approach to Coulomb spacecraft formation control. 
\end{abstract}
\section{Introduction}
The lifespan of a spacecraft formation is limited by the amount of propellant that can be brought onboard to be used for stationkeeping and other formation control maneuvers. As a measure to reduce propellant requirements for spacecraft formation control, 
Schuab {\it et al.} \cite{Prospects} first proposed the prospect of using the inter-spacecraft Coulomb forces for formation control in high orbits. By controlling the charge of the spacecraft in a formation, the Coulomb forces acting between the spacecraft can be controlled and this can, therefore, be used for formation control. Altering the charge of a spacecraft can be accomplished using negligible amounts of propellant. This, therefore, opens the possibility of spacecraft formation control using negligible amounts of propellant. 

With the benefit of nearly propellantless formation control comes some technical challenges inherent in Coulomb spacecraft formation control. There are two main difficulties: the first is the underactuated nature of Coulomb forces. The Coulomb forces act only along the line of sight between two charged spacecraft and come in action-reaction pairs. The Coulomb forces being internal forces cannot affect the angular momentum of the formation. In general, the use of conventional thrusters (or other external forces such as gravitational forces) in tandem with Coulomb actuation will be necessary to produce forces that cannot be achieved by Coulomb actuation alone \cite{ArunHybrid,Seo}. In this paper, the formations are constrained to be collinear because of this difficulty, and instead the focus of this paper is on addressing the second difficulty which is the nonlinear nature of the inputs. 

Recall that the Coulomb force is directly proportional to the product of the two charges. This means that in Coulomb spacecraft formation control, the equations of motion will be nonlinear in the the input, i.e. the charges. Most results in Coulomb formation control focus on the case of two spacecraft formations \cite{Prospects,ArunHybrid,Seo,Arun}. In the two spacecraft case, the nonlinearity with respect to the input can be simplified by redefining the input as a product of the two charges. The dynamics are then linear in the charge product and it is trivial to compute two individual charges that form a given charge product. For formations with more than two spacecraft, a simple redefinition of the variables in terms of products of charges cannot be done because constraints need to be added to ensure that a given combination of charge products is realizable. For example, in a three spacecraft formation it is impossible for all three charge products to be negative. Moreover, the magnitudes of the charge products are inter-dependent. 

Generally, there are two approaches that have been taken for Coulomb spacecraft formation control with formations of more than two spacecraft: the switching approach and the analytical approach. The switching approach exploits the fact that it is simpler to design controllers for Coulomb spacecraft formations with two spacecraft. The idea is to control two of the spacecraft at a time and then switch pairs according to some switching law \cite{wang,FELICETTI2016455}. The difficulty with this approach is that the switching law is very difficult to design and it becomes increasingly difficult as the number of spacecraft in the formation increases. It is then difficult to predict and overcome undesirable behaviors such as Zeno switching which can frequently arise when using state-dependent switching \cite{HybridSystems}.

The analytical approach involves deriving an analytical expression of the feedback control law. To simplify the dynamics to design a control law, constraints on the symmetry, shape, and charges will often be made. These constraints will often limit the scalability or applicability of the analytical control laws. Hussein and Schaub \cite{HUSSEIN2009738} impose symmetry constraints on collinear three spacecraft formations which enable them to reduce the number of inputs by then also constraining two of the spacecraft to have equal charges at all times. In Jones and Schuab \cite{Jones2014a}, a control law is derived by linearizing about a particular equilibrium and considering small deviations of the input from the equilibrium input. Their controller is then derived using linear control techniques. In Tahir and Narang-Siddarth \cite{CoulombNonlin}, a control law is derived for a collinear three spacecraft formation without performing any linearization; however, the resulting control law is a large symbolic expression and its use of dynamic inversions makes the control law difficult to scale for larger formations and nonrobust.  For four spacecraft formations, Vasavada and Schaub \cite{SolnCoul4} constrain the formations to be square formations. Lastly, Pettazzi {\it et al.} \cite{HybridCoulombIzzo}, derive a control law by disregarding the input nonlinearity and then perform a projection.



The approach in this paper for the stabilization of spacecraft formations of three spacecraft or more is unique from the state-of-the-art in that it is a completely optimization-based feedback control scheme. The approach addresses the nonlinearity with respect to the input without any constraints of symmetry or shape beyond collinearity. The approach in this paper is scalable in the sense that the optimization programs derived are independent of the number of spacecraft in the formation--the only difference with larger formations is that the size of the optimization variables and constraints will increase. Model predictive control (MPC) has received a lot of attention in aerospace applications in recent years (see the surveys \cite{7330727,doi:10.2514/1.G002507,MALYUTA2021282}); however, there has been no attention given to  Coulomb spacecraft formation control using MPC. This paper is the first step to filling in that gap.

The paper is organized as follows: \S II states the problem to be solved formally and writes out the dynamics of the Coulomb spacecraft formation. \S III derives the MPC algorithm by discretizing the dynamics, stating a nonconvex finite horizon optimization problem, and then deriving a convex relaxation of the  finite horizon optimization problem. \S IV provides a simulation of a four spacecraft formation and discusses the resulting control behavior as well as the computational burden of the MPC algorithm. \S V concludes the paper. 
\subsection{Notation}
Let $\mathbb{R}$ and $\mathbb{Z}$ denote the sets of real numbers and integers, respectively.  The positive real numbers and integers are denoted by $\mathbb{R}_+$ and $\mathbb{Z}_+$, respectively. The transpose is denoted using a superscript $\top$. The trace of a matrix $M$ is denoted by $\tr(M)$ and its rank is denoted by $\rank(M)$. A matrix $M$ is positive semidefinite if it symmetric and its eigenvalues are nonnegative, and $M\succeq 0$ denotes that $M$ is positive semidefinite. The $n\times n$ identity matrix is denoted by $I_n$ and the vector of dimension $n$ of all ones is denoted by $\onesvec_n$. If $x$ and $y$ are vectors of the same dimension, then $x\ge y$ denotes elementwise inequalities. 
\section{Problem Statement}
Consider a collinear spacecraft formation consisting of $N_s$ spacecraft. Let $x_1,x_2,\dots, x_{N_s}\in\mathbb{R}$ denote the position of each spacecraft and let $m_1,m_2,\dots, m_{N_s}\in\mathbb{R}_{+}$ denote the mass of each spacecraft. Each spacecraft has a charge $q_i\in\mathbb{R}$ which can be changed by an active charge control system onboard the spacecraft. 
\begin{figure}[h]
\centerline{\includegraphics[width=0.99\columnwidth]{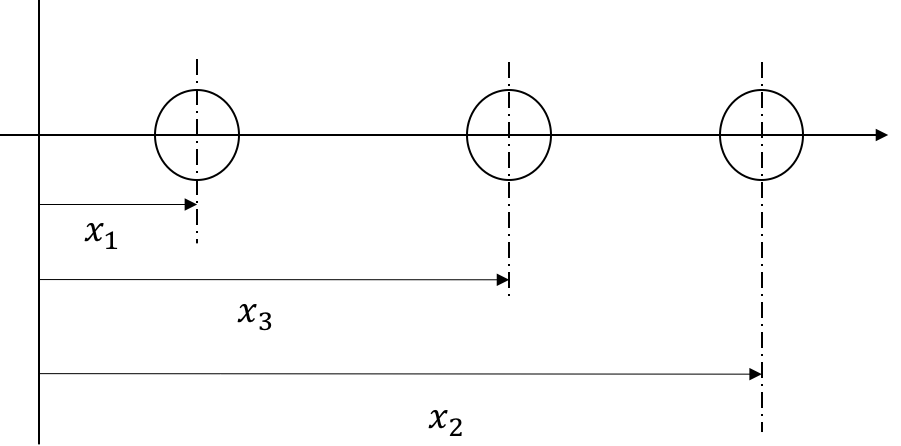}}
\caption{Depiction of a collinear spacecraft formation.}
\label{f:mm}
\end{figure}

The dynamics of the spacecraft in the formation is Newton's second law with the forces coming from Coulomb's law:
\begin{align}
\ddot{x}_i(t) = \frac{\kappa_c}{m_i}\sum_{j=1,j\ne i}^{N_s}\frac{x_i(t)-x_j(t)}{\|x_i(t)-x_j(t)\|^3}q_i(t)q_j(t),\label{scalardyn}
\end{align}
for all $i=1,\dots, N_s$, where $\kappa_c=8.99$e$+05\frac{\text{N}\cdot\text{m}^2}{(10\text{milliC})^2}$ is Coulomb's constant\footnote{Coulomb's constant is typically reported as $\kappa_c=8.99$e$+09\frac{\text{N}\cdot\text{m}^2}{\text{C}^2}$; however, for numerical purposes it is scaled down by four orders of magnitude to have units Newton meters squared per 10 milliCoulombs squared. So the units of the charge in \eqref{scalardyn} are 10's of milliCoulombs.}. 

By defining $x=[x_1,x_2,\dots,x_{N_s}]^\top$, \eqref{scalardyn} can be written compactly as:
\begin{align}
\ddot{x}(t) = \tilde{g}_c(x(t))L(q(t)),
\end{align}
where $\tilde{g}_c$ is a matrix which collects all of the inverse square terms along with the masses and Colulombs constant in \eqref{scalardyn} and $L$ is a function which collects all of the charge products that is defined as follows\footnote{The notation ``$L$" comes from the fact that $L$ can be constructed by vectorizing the lower-triangular portion of the matrix $qq^\top$.}:
\begin{align}
L(q) = [q_1q_2,\dots,q_1q_{N_s},q_2q_3,\dots,q_2q_{N_s},\dots,q_{N_s-1}q_{N_s}]^\top.\nonumber
\end{align}
So $L$ maps from $\mathbb{R}^{N_s}$ to $\mathbb{R}^m$ where $m={N_s \choose 2}$ is the number of unique pairs of spacecraft in the formation. Based on the definition of $L$, it will be useful to define a mapping from the index of $L$ to pairs of charges that are multiplied together in that index. Let the mapping be denoted by $\iota:\{1,\dots, m\}\to \{1,\dots,N_s\}\times \{1,\dots,N_s\}$. So, for example, if $N_s=3$, $\iota(1)=(1,2)$, $\iota(2)=(1,3)$, and $\iota(3)=(2,3)$.

Spacecraft formation control is primarily concerned with the control of the spacecraft relative positioning rather than the absolute positioning. Moreover, since Coulomb forces come in action-reaction pairs, only the relative positioning of the spacecraft can be controlled by Coulomb forces. The following relative coordinate system will be used:
\begin{align}
\xi_i=x_{i+1}-x_1,\forall i=1,\dots,N_s-1.\label{relativecoords}
\end{align}

The dynamics of the relative coordinates $\xi=[\xi_1,\xi_2,\dots,\xi_{N_s-1}]^\top$ can be written compactly as follows:
\begin{align}
\ddot{\xi}(t) = g_c(\xi(t))L(q(t))\label{compactsecondorder}
\end{align}
where the $i$th row of $g_c$ is computed by subtracting the first row of $\tilde{g}_c$ from the  $(i+1)$th row of $\tilde{g}_c$ and using the variable substitution in \eqref{relativecoords}. This can be written as a state-space model by defining $\nu=\dot{\xi}$ and the state variable
\begin{align*}
\Xi(t) = [\xi^\top(t),\nu^\top(t)]^\top
\end{align*}
as
\begin{align}
\dot{\Xi}(t) = \begin{bmatrix} 0 & I_{N_s-1} \\ 0 & 0  \end{bmatrix}\Xi(t)+\begin{bmatrix}0 \\  g_c(\xi(t)) \end{bmatrix}L(q(t)).\label{ssreldynamics}
\end{align}
Throughout this paper, it will be assumed that measurements of the full state of the system $\Xi(t)$ are always available for feedback control. 

In performing formation control, it may be desirable to impose constraints on the state $\Xi(t)$ and the input $q(t)$: 
\begin{align}
&\Xi_{\min}\le \Xi(t)\le \Xi_{\max}, \label{constrstate} \\
&q_{\min}\le q(t)\le q_{\max}, \label{constrinputs}
\end{align}
for all $t\ge 0$. The state constraints can be used to ensure two spacecraft do not get too close to each other to avoid potential collisions. The constraints on the charge are not necessarily used for fuel savings since the charge provides forces using negligible propellent. Rather, the input constraints may be necessary to keep the charge from getting too high which may damage some instrumentation onboard the spacecraft. 

The problem that is pursued in this paper is that of stabilizing a desired relative formation while ensuring that the constraints are met. This is stated more formally in the following:
\begin{problem}
Consider a given desired relative formation $\xi^{des}\in\mathbb{R}^{N_s-1}$. Using full-state charge feedback control, stabilize the desired states $\Xi^{des}=[{\xi^{des}}^\top,0^\top]^\top$ of the relative formation dynamics \eqref{ssreldynamics} while  ensuring the constraints \eqref{constrstate} and \eqref{constrinputs} on the state and inputs, respectively.  
\end{problem}
\section{MPC Algorithm}
The approach taken in this paper is an MPC approach which involves solving finite horizon optimal control problems at uniformly spaced sampling instances. To be implementable, the finite horizon problems need to be computationally tractable. This is done by first stating a finite horizon problem which is nonconvex and then finding a convex relaxation of that problem that can be solved efficiently using a convex optimization solver. 

\subsection{Discretization of the Dynamics}
The first step of the approach is to create the model that is used in solving the finite horizon problem computationally by an optimization solver. This is done by discretizing the dynamics \eqref{ssreldynamics}. Let $h$ denote the sampling period and let the square bracket notation denote the variable at each sampling stance, i.e. $(\cdot)[k] = (\cdot)(kh),$ for all $k=0,1,2,\dots$.

The input $q$ is implemented in a zeroth-order hold (ZOH) fashion, so 
\begin{align}
q(t)=q[k], \forall t\in[kh,(k+1)h).\label{e:ZOH}
\end{align}
Using a Runge-Kutta discretization scheme, an approximation of \eqref{ssreldynamics} with the input implemented using a ZOH as a discrete-time system of the form:
\begin{align}
\Xi[k+1] = A\Xi[k]+g_d(\xi[k])L(q[k]),\label{e:discretedynamics}
\end{align}
follows, where $g_d(\xi[k])$ is a nonlinear function and 
\begin{align*}
A = \begin{bmatrix} I_{N_s-1} & hI_{N_s-1}\\ 0 & I_{N_s-1}\end{bmatrix}.
\end{align*}

The finite horizon MPC algorithm will rely on convex optimization for fast computation. The discrete-time dynamics \eqref{e:discretedynamics} are nonconvex due to the nonlinear functions $g_d$ and $L$. To handle the nonconvexity of $g_d$, it will be assumed that the initial formation is close enough to the desired formation that $g_d(\xi[k])\approx g_d(\xi^{des})$ for all $k=0,1,2,\dots$.  The nonconvexity of $L$ will be handled later, but it will be useful to introduce a new variable $u:\mathbb{Z}\to \mathbb{R}^m$, which is the charge products. That is,
\begin{align}
u[k] = L(q[k]),
\end{align}
for all $k=0,1,2,\dots$.

Hence, the model to be used in the optimization problem can be written as:
\begin{align}
\Xi[k+1] = A\Xi[k]+Bu[k],\label{e:pldiscretedynamics}
\end{align}
where $B=g_c(\xi^{des})$. The dynamics expressed in \eqref{e:pldiscretedynamics} are affine in $\Xi$ and $u$ and therefore convex in $\Xi$ and $u$.

\subsection{Nonconvex MPC Optimization Problem} 
Now that the dynamics have been specified, a first cut of the finite horizon problem can be introduced. Let $N\in\mathbb{Z}_+$ be the prediction horizon. At each sampling instance, $k$, the full state $\Xi[k]$ is measured. Using this information, consider the following  finite horizon problem to be solved\footnote{
There are some differences between how the problem \eqref{e:MPC} is set up and the `archetypical' MPC problem (e.g. (2) in \cite{7330727} for a spacecraft formation control application). Notably, \eqref{e:MPC} does not have a constraint for a terminal set which is controlled invariant and does not have a terminal cost which, when designed properly, are useful for proving stability and recursive feasibility. The nonlinearities of the problem make it difficult to design a controlled invariant set and terminal cost. 
}:
\begin{subequations}\label{e:MPC}
   \begin{align}
& {\text{min}} 
& & \sum_{j=1}^{N} l(\hat{\Xi}[j],\hat{u}[j-1])+\sum_{j=1}^{N-1}l_\Delta(\hat{u}[j],\hat{u}[j-1])\label{e:cost}\\
& \text{s.t.}
& & \hat{\Xi}[j+1] = A\hat{\Xi}[j]+B\hat{u}[j], j= 0,\dots N-1,\\
&& & \hat{\Xi}[0] = \Xi[k] ,\\
& & & \hat{u}[j] = L(\hat{q}[j]),j=0,\dots N-1, \label{e:lq} \\
&&& \Xi_{\min}\le \hat{\Xi}[j]\le \Xi_{\max},j=0,\dots N-1,\\
&&& q_{\min}\le \hat{q}[j]\le q_{\max},j=0,\dots N-1,\label{e:inputconstrq}
\end{align}
where  
\begin{align}
&l(\hat{\Xi},\hat{u}) = (\hat{\Xi}-\Xi^{des})^\top \mathcal{Q} (\hat{\Xi}-\Xi^{des}) + \hat{u}^\top \mathcal{R} \hat{u},\label{e:quadcost}\\
&l_\Delta(\hat{u},\hat{v}) = (\hat{u}-\hat{v})^\top \mathcal{R}_{\Delta}(\hat{u}-\hat{v}), \label{e:smoothing} 
\end{align}
\end{subequations}
$\mathcal{Q},\mathcal{R},\mathcal{R}_\Delta\succeq0$. The `hat' notation denotes that the variables are predicted using the model and are decision variables for the optimization. 

The first terms in the cost \eqref{e:cost}, \eqref{e:quadcost} are standard quadratic costs on the charge products and the deviation from the desired states. These serve to stabilize the desired states with minimum input. The second set of terms \eqref{e:cost}, \eqref{e:smoothing} are there to promote smooth variation of the charge products over time. 

Once \eqref{e:MPC} is solved, the control is implemented by taking the first step:
\begin{align}
q[k] = \hat{q}^\star[0]\label{MPCcontrol}
\end{align}
where the $\star$ notation denotes that $\hat{q}^\star[0]$ is an optimal value, and implemented in a ZOH fashion using \eqref{e:ZOH}. The process is repeated at every time step. 
 
The optimization problem \eqref{e:MPC} is difficult to solve computationally due to the nonconvex constraint \eqref{e:lq}. The sequel will discuss a convex relaxation of this problem.  

\subsection{Convex Relaxation}
Constructing a convex relaxation of \eqref{e:MPC} begins by expressing the constraint \eqref{e:lq} as a series of (nonconvex) quadratic constraints. Notice that the constraint:
\begin{align*}
u_l=q_iq_j
\end{align*}
can be rewritten as a quadratic equation:
\begin{align*}
u_l = q^\top \mathcal{L}_{(i,j)} q,
\end{align*}
where 
\begin{align*}
\mathcal{L}_{(i,j)} = \frac{1}{2}(E_{(i,j)}+E_{(j,i)})
\end{align*}
and $E_{(i,j)}$ is a matrix which is zero everywhere except for it's $(i,j)$th element which is equal to 1. Note that the matrices $\mathcal{L}_{(i,j)}$ are indefinite.

Recall the mapping $\iota:\{1,\dots, m\}\to \{1,\dots,N_s\}\times \{1,\dots,N_s\}$, which was defined earlier when the function $L$ was defined. Using this notation, \eqref{e:lq} is equivalent to:
\begin{align}
\hat{u}_i[j] = \hat{q}[j]^\top\mathcal{L}_{\iota(i)}\hat{q}[j], \label{e:QC}
\end{align}
 for all $i=1,\dots, m$, and $j=0,\dots,N-1$. 
Therefore, \eqref{e:MPC} is equivalent to a nonconvex quadratically-constrained quadratic program (QCQP) with $N\times m$ quadratic constraints. 

One of the most popular methods of solving nonconvex QCQP's is to use the semidefinite relaxation (SDR) approach \cite{5447068}. This is done by first rewriting the right-hand side of \eqref{e:QC} as the follows:
\begin{align*}
\hat{q}[j]^\top\mathcal{L}_{\iota(i)}\hat{q}[j]= \tr\left(\mathcal{L}_{\iota(i)}\hat{q}[j]\hat{q}[j]^\top\right).
\end{align*}
This allows for the definition of new matrix variables 
\begin{align}
\hat{Q}[j]=\hat{q}[j]\hat{q}[j]^\top\label{e:redef}
\end{align}
for $ j=0,\dots, N-1$. By definition, $\hat{Q}[j]\succeq 0$ and $\rank(\hat{Q}[j])=1$. Therefore, \eqref{e:lq} can be expressed (with the change of variables) as the following:
\begin{subequations}\label{e:SDR}
\begin{align}
&\hat{u}_i[j] = \tr\left(\mathcal{L}_{\iota(i)}\hat{Q}[j]\right),\label{e:quadconstr}\\
&\hat{Q}[j]\succeq 0, \label{e:semidef}\\
&\rank(\hat{Q}[j])=1,\label{e:rank}
\end{align}
\end{subequations}
for all $i=1,\dots, m$ and $j=0,\dots, N-1$. Everything in \eqref{e:SDR} is convex except for the rank constraint \eqref{e:rank}. The SDR approach is to simply ignore the rank constraint. In this way, an upper bound on the cost of the nonconvex QCQP is achieved \cite{5447068,park2017generalheuristicsnonconvexquadratically}.

With the redefinition of variables in \eqref{e:redef}, the individual charges $\hat{q}[j]$ no longer appear in the optimization problem. So the input constraints \eqref{e:inputconstrq} cannot be directly applied. This can be mitigated by replacing the constraint on the charges with constraints on the charge products, i.e.  $u_{\min}\le \hat{u}[j]\le u_{\max}$ for $j=0,\dots N-1.$  Constraining the charge products rather than the charges themselves may still yield charges which violate the original constraints if, for instance, one charge is too large and the rest are small enough that the products fit within the new constraints. This is a tradeoff that is made for solvability.

Using the redefinition of variables and the semidefinite relaxation of the nonconvex quadratic constraints, the following is a convex relaxation of the problem \eqref{e:MPC}:
\begin{subequations}\label{e:MPC2}
   \begin{align}
& {\text{min}} 
& & \sum_{j=1}^{N} l(\hat{\Xi}[j],\hat{u}[j-1])+\sum_{j=1}^{N-1}l_\Delta(\hat{u}[j],\hat{u}[j-1])\nonumber\\
& & &+l_q\sum_{j=0}^{N-1}\tr(\hat{Q}[j]) \label{e:tracepen}\\ 
& \text{s.t.}
& & \hat{\Xi}[j+1] = A\hat{\Xi}[j]+B\hat{u}[j], j= 0,\dots N-1,\\
&& & \hat{\Xi}[0] = \Xi[k] ,\\
& & & \eqref{e:quadconstr},i=1,\dots, m, j=0,\dots,N-1,\\
&&& \hat{Q}[j]\succeq 0, j=0,\dots,N-1,\\
&&& \Xi_{\min}\le \hat{\Xi}[j]\le \Xi_{\max},j=0,\dots N-1,\\
&&& u_{\min}\le \hat{u}[j]\le u_{\max},j=0,\dots N-1.
\label{e:optimalMPCprob}
\end{align}
\end{subequations}

The question remains of how to find $\hat{q}^\star[0]$ to be implemented as the control \eqref{MPCcontrol} once \eqref{e:MPC2} is solved. From the solution $\hat{Q}^\star[0]$, the charge $\hat{q}^\star[0]$ can be recovered by finding $\hat{q}^\star[0]$ such that $\hat{q}^\star[0]\hat{q}^\star[0]^\top$ is closest to $\hat{Q}^\star[0]$. Let the eigenvalues of $\hat{Q}^\star[0]$ be denoted $\lambda_1,\dots,\lambda_{N_s}$ where $0\le\lambda_1\le \dots\le \lambda_{N_s}$ with associated eigenvectors $v_1,\dots,v_{N_s}$. In the sense of minimizing the Frobenius norm of the difference (i.e. the sum of squares of elements of $\hat{q}^\star[0]\hat{q}^\star[0]^\top-\hat{Q}^\star[0]$) the best charge $\hat{q}^\star[0]$ is the following \cite{5447068},\cite[\S 7.4.2]{HornyJohnson}:
\begin{align}
\hat{q}^\star[0]=\sqrt{\lambda_{N_s}}v_{N_s}.\label{e:chargesfrommatrix}
\end{align}
Notice that if $\hat{Q}^\star[0]$ is rank one, then using \eqref{e:chargesfrommatrix} yields $\hat{Q}^\star[0]=\hat{q}^\star[0]\hat{q}^\star[0]^\top$. To ensure that the matrices $\hat{Q}[j]$ are as close to rank one as possible, a trace penalty is added to the cost function \eqref{e:tracepen}, where $l_q\ge 0$ is a weighting term. The trace penalty promotes solutions with low rank similarly to how $\ell_1$ penalties are used to promote solutions that are sparse \cite{nuclearnorm}.
\subsection{Summary of the Algorithm}
To summarize, the MPC algorithm is the following:
\begin{enumerate}[(1)]
\item At timestep $k$, measure the full state of the system $\Xi[k]$.
\item Solve the convex SDR problem \eqref{e:MPC2}.
\item Compute an eigendecomposition of $\hat{Q}^\star[0]$. 
\item Compute the charges $\hat{q}^\star[0]$ using \eqref{e:chargesfrommatrix}.
\item Implement the charges using \eqref{e:ZOH} and \eqref{MPCcontrol}.
\item Wait until the next timestep and repeat. 
\end{enumerate}

\section{Simulation}

In this example, a four spacecraft collinear formation is considered. The desired relative formation is the following:
\begin{align*}
\xi^{des} = [50.0,100.0,150.0]^\top,
\end{align*}
and the initial condition is the following:
\begin{align*}
\Xi[0] = [53.0,109.0,147.0,0.0,0.0,0.0]^\top.
\end{align*}
The sampling period is chosen to be $h=0.5$ seconds.

The following are the parameters chosen for the convex SDR problem \eqref{e:MPC2}: $N=9, \mathcal{Q} = \diag(I_3, 20^2I_3), \mathcal{R}= 0,\mathcal{R}_\Delta = 10^8I_6, l_q=1.5,$ and the state constraints are:
\begin{align*}
\Xi_{\min} = \Xi^{des}-10\cdot\onesvec_{2(N_s-1)}, \Xi_{\max} = \Xi^{des}+10\cdot\onesvec_{2(N_s-1)}.
\end{align*}

No constraints on the charge products were implemented. Instead, constraints on the charge were implemented by saturating the value derived from the eigenvalue decomposition \eqref{e:chargesfrommatrix} such that the maximum magnitude of the charge is 1milliC. 

\begin{figure}[h]
\centerline{\includegraphics[width=0.99\columnwidth]{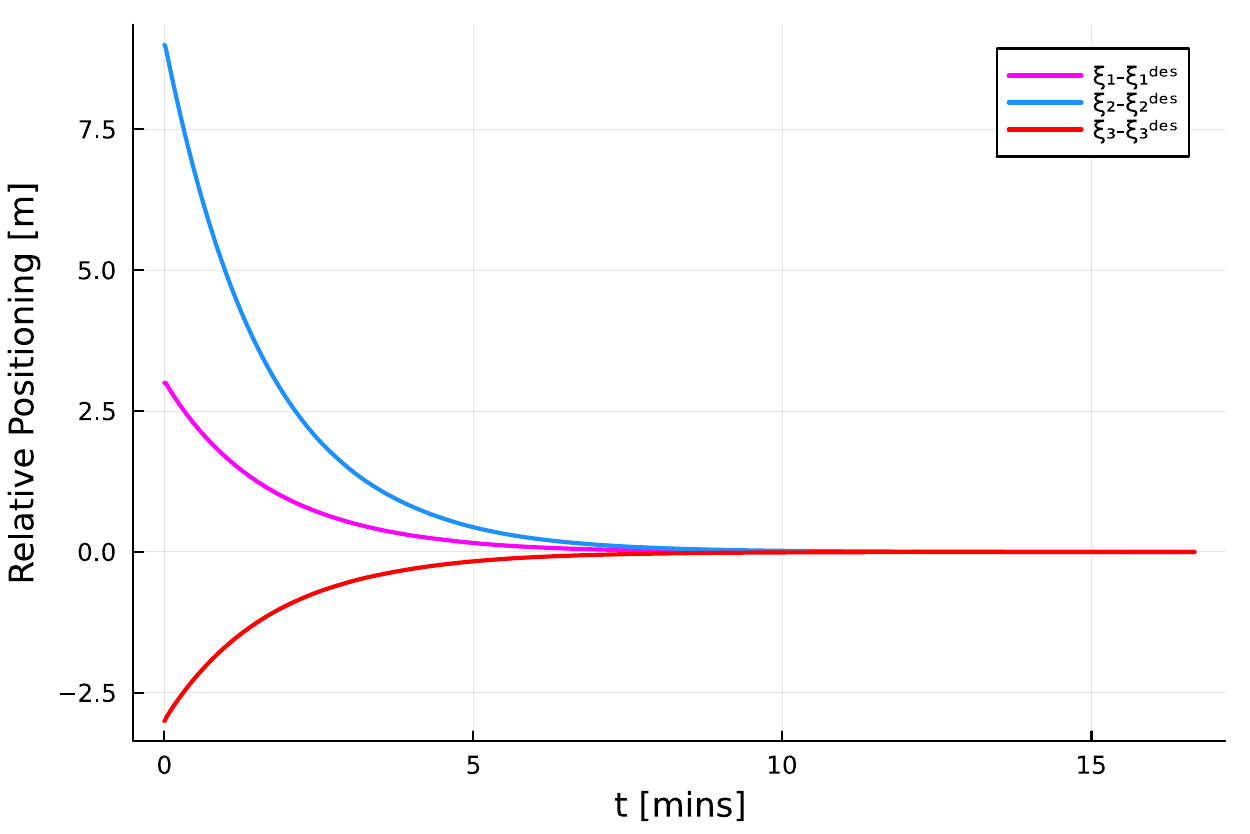}}
\caption{Trajectory of the deviation of the relative positions from the desired states. }
\label{f:stabilize}
\end{figure}

\begin{figure}[h]
\centerline{\includegraphics[width=0.99\columnwidth]{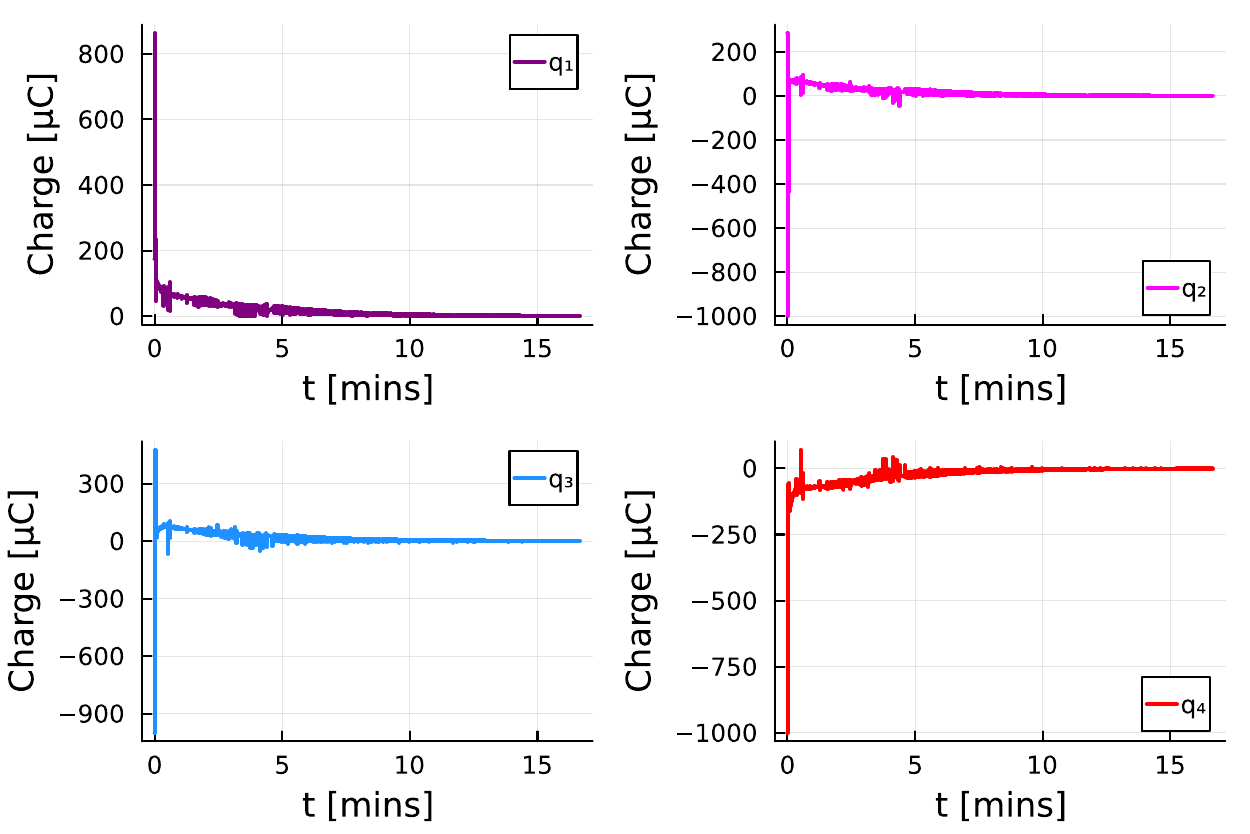}}
\caption{Charges computed to achieve the trajectory in Fig. \ref{f:stabilize}. }
\label{f:charges}
\end{figure}

The simulation and optimization was performed in \verb|Julia|. The discretization and linearization was performed using \verb|RobotDynamics.jl|\footnote{https://rexlab.ri.cmu.edu/RobotDynamics.jl/stable/}. The convex optimizer solver will be discussed at the end of this section. Fig. \ref{f:stabilize} shows that the relative positions are stabilized to the desired states, and the sequence of charges that are computed using the MPC algorithm to stabilize the desired states are shown in Fig. \ref{f:charges}.

\begin{figure}[h]
\centerline{\includegraphics[width=0.99\columnwidth]{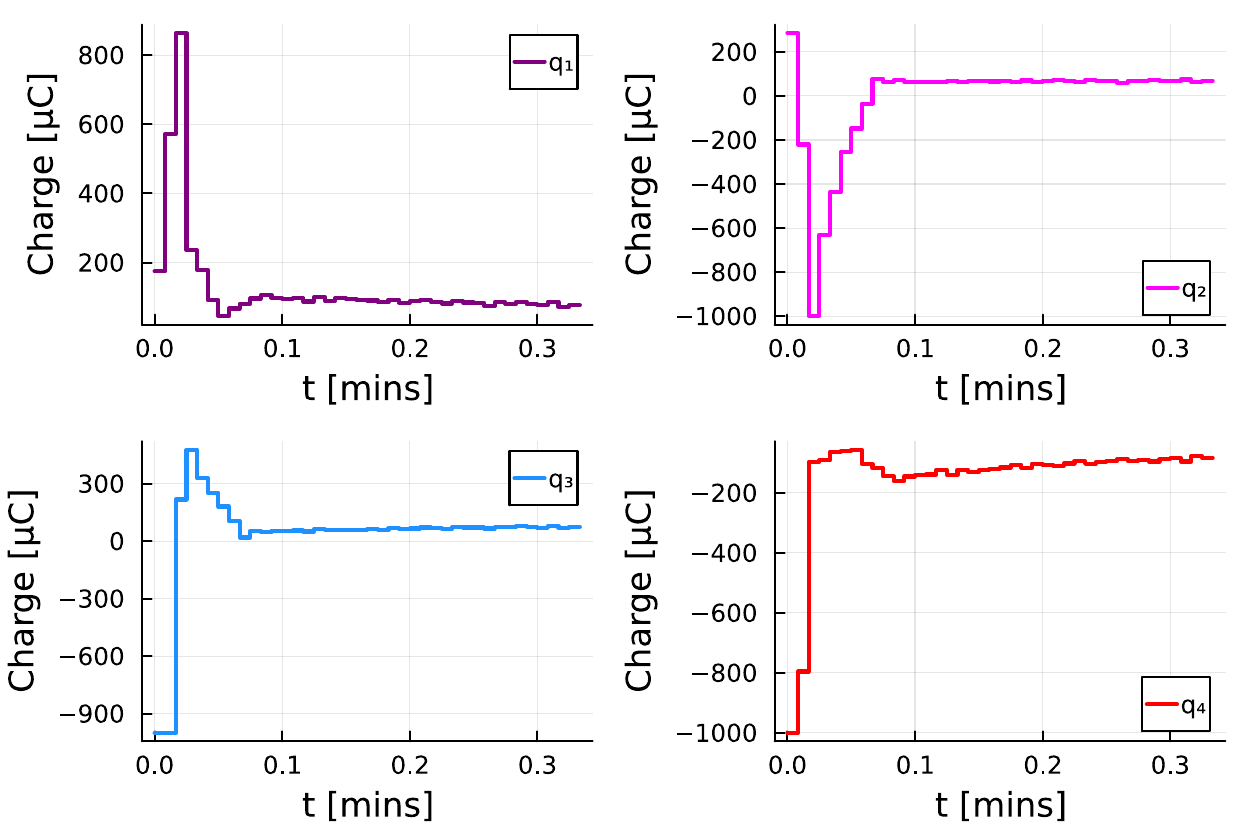}}
\caption{The first few seconds of the charges in Fig \ref{f:charges}.}
\label{f:zoom1}
\end{figure}

There are some interesting behaviors in the charge inputs that can be observed by zooming into different slices of Fig. \ref{f:charges}. Firstly, consider Fig. \ref{f:zoom1} which shows that first few seconds of the charge input. The largest charges occur at the very beginning of the simulation. In fact, a couple of the charges are saturated by the 1milliC limit. Since the spacecraft start from rest, the large charges constitute an initial jerk to get moving in the right direction and then the charges settle. 

\begin{figure}[h]
\centerline{\includegraphics[width=0.99\columnwidth]{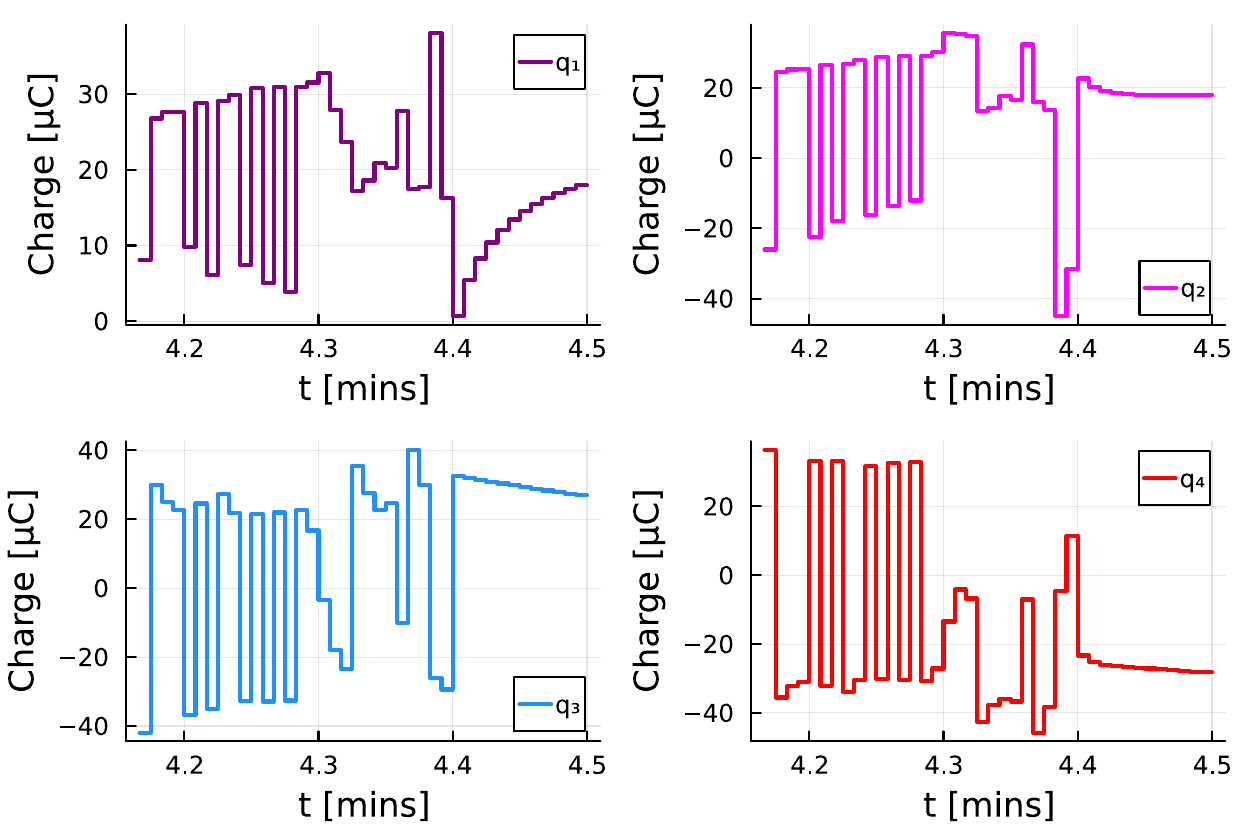}}
\caption{A close up showing the alternating `push-pull' behavior.}
\label{f:zoom2}
\end{figure}

As the spacecraft get closer to the desired states, the charges appear to have a noisy pattern in Fig. \ref{f:charges}. By zooming into this apparent noise, some interesting behavior is observed which is shown in Fig. \ref{f:zoom2}. In this plot it can be seen that $q_1$ is always positive and the other three alternate between having positive and negative charges. This means that the spacecraft are alternating between pushing and pulling each other. This is likely due to the fact that as the desired states are being approached, there is a need to keep moving towards the goal and at the same time begin to slow down. There is no combination of charges that can be found such that all spacecraft pull towards each other since a pulling force requires opposite charges and not all spacecraft can have opposite charges. Hence the need for alternating between pushing and pulling. This illustrates why predictive control is a good approach for Coulomb spacecraft control.

Different trajectories can be obtained by varying the parameters. In the above simulation, the penalty on the relative velocities in $\mathcal{Q}$ is high which increases the time required to stabilize the desired states and also decreases the charges compared to a $\mathcal{Q}$ with lower penalties on the relative velocities. 
\subsection{Discussion of Computational Burden}
The example is solved using the \verb|SCS| solver \cite{ocpb:16} through \verb|Convex.jl| \cite{convexjl} with solver warmstarting from the solution at the previous timestep. Moreover, \verb|SparseArrays.jl|\footnote{https://sparsearrays.juliasparse.org/dev/} was used to account for the sparsity patterns in the $A,B,\mathcal{L}_{(i,j)}$ matrices when setting up the problem. The simulation was done on a Macbook Pro laptop with an Intel Core i5 at 2.4 GHz. In the simulation shown above, the convex problem \eqref{e:MPC2} would take around 0.1 seconds or less to converge at each sampling instance. 

Further study should be devoted to the computational burden. It is possible that customizing the solver (by exploiting the problem structure in the numerical scheme) can potentially yield several orders of magnitude of improved convergence speed (cf. \cite{Dueri}).  Other approaches to solving nonconvex QCQP's may be much faster. For example, using an ADMM-based approach allows for parallelization \cite{7517329}. 

\section{Conclusions}
An MPC approach to stabilizing collinear Coulomb spacecraft formations to desired configurations was derived in this paper. The difficulty with Coulomb spacecraft formation control is the nonlinearity with respect to the input. This is addressed by stating the finite horizon optimization problem as a nonconvex QCQP which can be solved approximately using a convex relaxation that can be solved sufficiently fast. A simulation of a four spacecraft formation was provided to show that the MPC scheme derived is stabilizing. Interesting behaviors were observed in the solution such as the alternating `push-pull' behavior. This need to alternate between pushing and pulling illustrates why optimizing over a prediction horizon is a prudent approach to Coulomb spacecraft formation control. It also suggests that parameterizing the input using periodic functions (e.g. sinusoids, square waves) could be beneficial for other control schemes.

\bibliography{MLbib}

\begin{thebibliography}{10}
\providecommand{\url}[1]{#1}
\csname url@samestyle\endcsname
\providecommand{\newblock}{\relax}
\providecommand{\bibinfo}[2]{#2}
\providecommand{\BIBentrySTDinterwordspacing}{\spaceskip=0pt\relax}
\providecommand{\BIBentryALTinterwordstretchfactor}{4}
\providecommand{\BIBentryALTinterwordspacing}{\spaceskip=\fontdimen2\font plus
\BIBentryALTinterwordstretchfactor\fontdimen3\font minus
  \fontdimen4\font\relax}
\providecommand{\BIBforeignlanguage}[2]{{%
\expandafter\ifx\csname l@#1\endcsname\relax
\typeout{** WARNING: IEEEtran.bst: No hyphenation pattern has been}%
\typeout{** loaded for the language `#1'. Using the pattern for}%
\typeout{** the default language instead.}%
\else
\language=\csname l@#1\endcsname
\fi
#2}}
\providecommand{\BIBdecl}{\relax}
\BIBdecl

\bibitem{Prospects}
H.~Schaub, G.~G. Parker, and L.~B. King, ``Challenges and prospects of
  {Coulomb} spacecraft formation control,'' \emph{Journal of the Astronautical
  Sciences}, vol.~52, no. 1-2, pp. 169--193, 2004.

\bibitem{invariantshape}
I.~I. Hussein and H.~Schaub, ``Invariant shape solutions of the spinning three
  craft {C}oulomb tether problem,'' \emph{Celestial Mechanics and Dynamical
  Astronomy}, vol.~96, pp. 137--157, 2006.

\bibitem{Berryman}
J.~Berryman and H.~Schaub, ``Analytical charge analysis for 2- and 3-craft
  {C}oulomb formations,,'' \emph{Journal of Guidance, Control, and Dynamics},
  vol.~30, no.~6, pp. 1701--1710, 2007.

\bibitem{Jones2014a}
D.~R. Jones and H.~Schaub, ``Collinear three-craft {Coulomb} formation
  stability analysis and control,'' \emph{Journal of Guidance, Control, and
  Dynamics}, vol.~37, no.~1, pp. 224--232, 2014.

\bibitem{CoulombNonlin}
A.~M. Tahir and A.~{Narang-Siddarth}, ``Constructive nonlinear approach to
  {C}oulomb formation control,'' in \emph{Proceedings of AIAA Guidance,
  Navigation, and Control Conference}, Kissimmee, FL, 2018.

\bibitem{wang}
S.~Wang and H.~Schaub, ``Coulomb control of non-equilibrium fixed shape
  triangular three-vehicle cluster,'' \emph{Journal of Guidance, Control, and
  Dynamics}, vol.~34, no.~1, pp. 259--270, 2011.

\bibitem{ArunHybrid}
A.~Natarajan and H.~Schaub, ``Hybrid control of orbit normal and along-track
  two-craft coulomb tethers,,'' \emph{Aerospace Science and Technology},
  vol.~13, pp. 183--191, 2009.

\bibitem{HybridCoulombIzzo}
L.~Pettazzi, H.~Kr\"{u}ger, S.~Theil, and D.~Izzo, ``Electrostatic force for
  swarm navigation and reconfiguration,'' \emph{Acta Futura}, vol.~3, pp.
  80--86, 2009.

\bibitem{Seo}
M.~W. Memon, M.~Nazari, D.~Seo, and E.~A. Butcher, ``Fuel efficiency of fully
  and underconstrained {Coulomb} formations in slightly elliptic reference
  orbits,'' \emph{IEEE Transactions on Aerospace and Electronic Systems},
  vol.~57, no.~6, pp. 4171--4187, 2021.

\bibitem{Allocation}
T.~A. Johansen and T.~I. Fossen, ``Control allocation--a survey,''
  \emph{Automatica}, vol.~49, pp. 1087--1103, 2013.

\bibitem{BilinearEquations}
C.~R. Johnson, H.~Smigoc, and D.~Yang, ``Solution theory for systems of
  bilinear equations,'' \emph{Linear and Multilinear Algebra}, vol.~61, no.~12,
  pp. 1553--1566, 2013.

\bibitem{PSD_fixed_rank}
B.~Vandereycken, P.-A. Absil, and S.~Vandewalle, ``{Embedded geometry of the
  set of symmetric positive semidefinite matrices of fixed rank},'' in
  \emph{{2009 IEEE/SP 15th Workshop on Statistical Signal Processing}},
  Cardiff, UK, 2009, pp. {389--392}.

\bibitem{zheng2023riemannian}
S.~Zheng, W.~Huang, B.~Vandereycken, and X.~Zhang, ``{Riemannian optimization
  using three different metrics for Hermitian PSD fixed-rank constraints: an
  extended version},'' arXiv:2204.07830, 2023.

\bibitem{AMS_manifolds}
P.-A. Absil, R.~Mahony, and R.~Sepulchre, \emph{Optimization Algorithms on
  Matrix Manifolds}.\hskip 1em plus 0.5em minus 0.4em\relax Princeton
  University Press, 2008.

\bibitem{boumal2023intromanifolds}
N.~Boumal, \emph{An Introduction to Optimization on Smooth Manifolds}.\hskip
  1em plus 0.5em minus 0.4em\relax Cambridge University Press, 2023.

\bibitem{Pymanopt}
J.~Townsend, N.~Koep, and S.~Weichwald, ``Pymanopt: A {P}ython toolbox for
  optimization on manifolds using automatic differentiation,'' \emph{Journal of
  Machine Learning Research}, vol.~17, no. 137, pp. 1--5, 2016.

\bibitem{CoulombForce}
C.~R. Seubert, L.~A. Stiles, and H.~Schaub, ``Effective {C}oulomb force
  modeling for spacecraft in earth orbit plasmas,'' \emph{Advances in Space
  Research}, vol.~54, no.~2, pp. 209--220, 2014.

\end{thebibliography}


\begin{thebibliography}{10}
\providecommand{\url}[1]{#1}
\csname url@samestyle\endcsname
\providecommand{\newblock}{\relax}
\providecommand{\bibinfo}[2]{#2}
\providecommand{\BIBentrySTDinterwordspacing}{\spaceskip=0pt\relax}
\providecommand{\BIBentryALTinterwordstretchfactor}{4}
\providecommand{\BIBentryALTinterwordspacing}{\spaceskip=\fontdimen2\font plus
\BIBentryALTinterwordstretchfactor\fontdimen3\font minus
  \fontdimen4\font\relax}
\providecommand{\BIBforeignlanguage}[2]{{%
\expandafter\ifx\csname l@#1\endcsname\relax
\typeout{** WARNING: IEEEtran.bst: No hyphenation pattern has been}%
\typeout{** loaded for the language `#1'. Using the pattern for}%
\typeout{** the default language instead.}%
\else
\language=\csname l@#1\endcsname
\fi
#2}}
\providecommand{\BIBdecl}{\relax}
\BIBdecl

\bibitem{Prospects}
H.~Schaub, G.~G. Parker, and L.~B. King, ``Challenges and prospects of
  {Coulomb} spacecraft formation control,'' \emph{Journal of the Astronautical
  Sciences}, vol.~52, no. 1-2, pp. 169--193, 2004.

\bibitem{ArunHybrid}
A.~Natarajan and H.~Schaub, ``Hybrid control of orbit normal and along-track
  two-craft {C}oulomb tethers,'' \emph{Aerospace Science and Technology},
  vol.~13, pp. 183--191, 2009.

\bibitem{Seo}
M.~W. Memon, M.~Nazari, D.~Seo, and E.~A. Butcher, ``Fuel efficiency of fully
  and underconstrained {Coulomb} formations in slightly elliptic reference
  orbits,'' \emph{IEEE Transactions on Aerospace and Electronic Systems},
  vol.~57, no.~6, pp. 4171--4187, 2021.

\bibitem{Arun}
A.~Natarajan and H.~Schaub, ``Linear dynamics and stability analysis of a
  two-craft {C}oulomb tether formation,'' \emph{Journal of Guidance, Control,
  and Dynamics}, vol.~29, no.~4, pp. 831--839, 2006.

\bibitem{wang}
S.~Wang and H.~Schaub, ``Coulomb control of non-equilibrium fixed shape
  triangular three-vehicle cluster,'' \emph{Journal of Guidance, Control, and
  Dynamics}, vol.~34, no.~1, pp. 259--270, 2011.

\bibitem{FELICETTI2016455}
L.~Felicetti and G.~B. Palmerini, ``Analytical and numerical investigations on
  spacecraft formation control by using electrostatic forces,'' \emph{Acta
  Astronautica}, vol. 123, pp. 455--469, 2016.

\bibitem{HybridSystems}
R.~Goebel, R.~G. Sanfelice, and A.~R. Teel, \emph{Hybrid Dynamical Systems:
  Modeling, Stability, and Robustness}.\hskip 1em plus 0.5em minus 0.4em\relax
  Princeton University Press, 2012.

\bibitem{HUSSEIN2009738}
I.~Hussein and H.~Schaub, ``Stability and control of relative equilibria for
  the three-spacecraft coulomb tether problem,'' \emph{Acta Astronautica},
  vol.~65, no.~5, pp. 738--754, 2009.

\bibitem{Jones2014a}
D.~R. Jones and H.~Schaub, ``Collinear three-craft {Coulomb} formation
  stability analysis and control,'' \emph{Journal of Guidance, Control, and
  Dynamics}, vol.~37, no.~1, pp. 224--232, 2014.

\bibitem{CoulombNonlin}
A.~M. Tahir and A.~{Narang-Siddarth}, ``Constructive nonlinear approach to
  {C}oulomb formation control,'' in \emph{Proceedings of AIAA Guidance,
  Navigation, and Control Conference}, Kissimmee, FL, 2018.

\bibitem{SolnCoul4}
H.~Vasavada and H.~Schaub, ``Analytic solutions for equal mass 4-craft static
  {C}oulomb formation,'' \emph{Journal of the Astronautical Sciences}, vol.~56,
  pp. 17--40, 2008.

\bibitem{HybridCoulombIzzo}
L.~Pettazzi, H.~Kr\"{u}ger, S.~Theil, and D.~Izzo, ``Electrostatic force for
  swarm navigation and reconfiguration,'' \emph{Acta Futura}, vol.~3, pp.
  80--86, 2009.

\bibitem{7330727}
E.~N. Hartley, ``A tutorial on model predictive control for spacecraft
  rendezvous,'' in \emph{Proceedings of the European Control Conference}, Linz,
  Austria, 2015, pp. 1355--1361.

\bibitem{doi:10.2514/1.G002507}
U.~Eren, A.~Prach, B.~B. Ko\c{c}er, S.~V. Rakovi\'{c}, E.~Kayacan, and
  B.~A\c{c}\i{}kme\c{s}e, ``Model predictive control in aerospace systems:
  Current state and opportunities,'' \emph{Journal of Guidance, Control, and
  Dynamics}, vol.~40, no.~7, pp. 1541--1566, 2017.

\bibitem{MALYUTA2021282}
D.~Malyuta, Y.~Yu, P.~Elango, and B.~A{\c c}ıkme{\c s}e, ``Advances in
  trajectory optimization for space vehicle control,'' \emph{Annual Reviews in
  Control}, vol.~52, pp. 282--315, 2021.

\bibitem{5447068}
Z.-Q. Luo, W.-K. Ma, A.~M.-C. So, Y.~Ye, and S.~Zhang, ``Semidefinite
  relaxation of quadratic optimization problems,'' \emph{IEEE Signal Processing
  Magazine}, vol.~27, no.~3, pp. 20--34, 2010.

\bibitem{park2017generalheuristicsnonconvexquadratically}
\BIBentryALTinterwordspacing
J.~Park and S.~Boyd, ``General heuristics for nonconvex quadratically
  constrained quadratic programming,'' 2017. [Online]. Available:
  \url{https://arxiv.org/abs/1703.07870}
\BIBentrySTDinterwordspacing

\bibitem{HornyJohnson}
R.~A. Horn and C.~R. Johnson, \emph{Matrix Analysis}, 2nd~ed.\hskip 1em plus
  0.5em minus 0.4em\relax Cambridge University Press, 2012.

\bibitem{nuclearnorm}
B.~Recht, M.~Fazel, and P.~A. Parrilo, ``Guaranteed minimum-rank solutions of
  linear matrix equations via nuclear norm minimization,'' \emph{SIAM Review},
  vol.~52, no.~3, pp. 471--501, 2010.

\bibitem{ocpb:16}
B.~O'Donoghue, E.~Chu, N.~Parikh, and S.~Boyd, ``Conic optimization via
  operator splitting and homogeneous self-dual embedding,'' \emph{Journal of
  Optimization Theory and Applications}, vol. 169, no.~3, pp. 1042--1068, June
  2016.

\bibitem{convexjl}
M.~Udell, K.~Mohan, D.~Zeng, J.~Hong, S.~Diamond, and S.~Boyd, ``Convex
  optimization in {J}ulia,'' \emph{SC14 Workshop on High Performance Technical
  Computing in Dynamic Languages}, 2014.

\bibitem{Dueri}
D.~Dueri, B.~A\c{c}\i{}kme\c{s}e, D.~P. Scharf, and M.~W. Harris, ``Customized
  real-time interior-point methods for onboard powered-descent guidance,''
  \emph{Journal of Guidance, Control, and Dynamics}, vol.~40, no.~2, pp.
  197--212, 2017.

\bibitem{7517329}
K.~Huang and N.~D. Sidiropoulos, ``Consensus-{ADMM} for general quadratically
  constrained quadratic programming,'' \emph{IEEE Transactions on Signal
  Processing}, vol.~64, no.~20, pp. 5297--5310, 2016.

\end{thebibliography}

\end{document}